# Detection of melting by *in-situ* observation of spherical-drop formation in laser-heated diamond-anvil cells

Thomas Pippinger [a], Leonid Dubrovinsky [b], Konstantin Glazyrin [b], Ronald Miletich [a),c)] and Natalia Dubrovinskaia [a),d)]

[a] FG Mineralphysik, Institut für Geowissenschaften, Universität Heidelberg, Heidelberg, Germany

[b] Bayerisches Geoinstitut, Universität Bayreuth, Bayreuth, Germany

[c] Institut für Mineralogie und Kristallographie, Universität Wien, Wien, Austria

[d] Lehrstuhl für Kristallographie, Physikalisches Institut, Universität Bayreuth, Bayreuth, Germany

Abstract:

A simple method for detection of melting event in laser-heated diamond anvil cells (DACs) is introduced. The melting is registered optically by the formation of spherical drops of the investigated material as heated in an inert pressure transmitting medium. Feasibility of the method is demonstrated on the examples of metal (iron and gold) and iron oxide ($Fe_2O_3$), materials molten at pressures over 40 GPa employing a portable laser heating system.

1. Introduction

Laser induced heating in diamond anvil cells is a common technique to achieve simultaneous high-pressure and high-temperature (HP-HT) conditions on a sample [1,2]. The powerful method allows simulating extreme conditions comparable to deep inner Earth environment i. e. ( pressures (P) above 100 GPa and temperatures (T) exceeding 3000K ) [3]. The possibility to increase temperature up to melting point of a sample under static high-pressure conditions is of interdisciplinary interest. Not only melting behavior of different materials can be investigated, but also studies of a temperature dependence of high-pressure phase transformations, changes in the order-disorder state or even in-situ synthesis under extreme conditions becomes possible. For *in-situ* investigations this technique has been coupled with different analytical methods, such as the powder or single-crystal X-ray diffraction and Raman spectroscopy [3,4] and becomes available on most of the extreme conditions beamlines at large scale facilities, such as synchrotron [5-7]. Although so far large stationary equipment dominated experimental setups, developments in laser-heating technique and optical systems lead to designing and application of more compact mobile devices, which can be used for multiple purposes [5,8,9,10].

Monitoring of sample's behavior during high-P,T experiments in DACs is essential and should be conducted carefully. It can be realized through optical observations, collecting diffraction patterns, X-ray absorption [10] or Raman spectra [4]. All mentioned methods are associated with specific challenges, which can lead to ambiguities in detecting a modification in the aggregate state of the material under investigation. For instance, optical observations are linked to a subjective experience of the observing person and in many cases only optically visible surface changes can be recognized. Identification of melting using X-ray diffraction is in general possible at high brilliance synchrotron facilities equipped with an appropriate setup [11,12]. The melting event is manifested in the diffraction data either by the disappearance of Bragg intensities and/or by the appearance of diffuse scattering due to the short-range order in the liquid state. In both cases the melt has to be stable for some time that might complicate experiments due to possible chemical reactions with the pressure medium or the diamond anvils [11,12]. Melting detection by X-ray absorption is also a synchrotron dependent technique and possesses the same problems like those mentioned for X-ray diffraction experiments. Here we present a method for the immediate detection of a melting event visually observable in laser-heated diamond anvil cells and demonstrate its experimental application on melting gold, iron and hematite ($Fe_2O_3$) samples under static high-pressure conditions.

2. Materials

Hematite single-crystals ($Fe_2O_3$) were synthesized in a platinum crucible by applying flux-based synthesis in the ternary system $B_2O_3 - Bi_2O_3 - Fe_2O_3$ [13]. The three agents (of a corresponding purity) were purchased form Sigma-Aldrich 99.98 %, Merk 99.0 %, Alfa Aesar 99.99 %, respectively. The prepared crystals were carefully cleaned and analyzed using the powder X-ray diffraction and single-crystal X-ray diffraction (using a Phillips XPert and a Huber 5042 diffractometer)). Iron (Fe, Goodfellow, 99.99+ %) and gold (Au, Goodfellow, 99.95 %) were used in form of pieces of commercially available foils with a size of 10-20 μm. Purity of materials was verified by the electron microprobe analysis and, additionally, for iron and hematite by Mössbauer spectroscopy (at Bayerisches Geoinstitut).

3. Experimental - Laser heating in DACs

Laser heating in our experiments was realized using the portable laser-heating system for diamond anvil cells described in detail in Ref. 5 and Ref. 8. Main parts of the system are the SPI 100 fiber laser (100 W, 1064 nm, pulsed or continuous (CW) mode) and the latest version of the universal laser-heating head (UniHead) [8]. Figure 1 shows the schematic of the experimental setup. The output of the SPI laser has a Gaussian power distribution with a diameter of ~3 mm at 1/e² and is focused by means of a π-shaper (MolTech GmbH) and UniHead optics to ~40 μm FWHM. The optical system forms a homogeneous intensity plateau at the sample position. The modified (with respect to the description in Ref. 8) visualization and optical output module consists of a μEYE$^{TM}$ CCD camera, a focusing lens, projecting a central part (about 10 μm in diameter) of the image of the heated spot on the end of the optical fiber, and a beam-splitter cube (50/50, Edmund Scientific Inc.) (Fig. 2). The beam-splitter cube is movable and can be fixed in two positions, one of them outside the optical path (and all light radiated by the heated spot transmits to the optical fiber) or within the optical path (in this position 50% of light is transferred to the CCD camera for visual observations).

3.1 Temperature measurements and calibration

For the temperature measurement we employed the multi-wavelenght spectroradiometry [5,10]. Thermal emission spectra were collected during laser heating with exposures of 0.05 to 1 second and a repetition rate of 0.1 to 10 Hz in different experiments. We fitted the measured spectra into Plank's radiation function [6,10]:

$I(\lambda) = c_1\ \varepsilon\ \lambda^{-5} / \exp(c_2 / (\lambda\ T)^{-1})$,

where I is the wavelength dependent intensity of the spectrum, $c_1 = 2\ \pi\ h\ c$, $c_2 = h\ c / k$, ε is the emissivity; T- the temperature; h- the Planck's constant; c- the speed of light; and k- the Boltzmann constant. Emissivity of a black body is equal to one, but for most materials emissivity is unknown. We work with the assumption that ε is linearly wavelength dependent. The typical uncertainty in the fitting procedure is within 10 K. In order to calibrate the system we melted 10, 15 and 20 μm thick Pt foils at ambient pressure. The foils were heated from one side by a gradual increase of the laser power, while radiation spectra were collected from the opposite until the metal melted and a hole formed. The melting temperature measured this way was not dependent on the thickness of the foil. However, if temperature was measured from the same side, from which foil was heated, the temperature of the hole formation depends on the thickness of the metal and overheating reaches 150 K (for 10 μm thick foil) to 300 K (for 25 μm thick foil). Consequently, in all melting experiments described below we report temperatures measured from the side where no overheating was observed during calibration procedure.

3.2 High pressure experiments

For compression of the samples we used DACs equipped with the anvils with the culet size of 250 μm to 390 μm for individual experiments [14]. Rhenium gaskets, pre-indented to about 30 to 70 μm in thickness and with holes of 125 to 160 μm initial diameter, were employed. Small pieces of Au or Fe of irregular shape with a maximum dimension of about 40 μm and a ruby pressure marker were loaded directly into the pressure chambers. The average estimated uncertainty in pressure measurements by means of ruby fluorescence is within 0.5 GPa for the given pressure range. No corrections on thermal pressure were introduced, but we speculate that possible thermal pressure (about 2 GPa at 2000 K) may be compensated by thermal relaxation of the sample near melting. Neon and argon were used as both a pressure transmitting medium and a thermal insulator, and loaded into the DACs at a pressure of 1.4 kbar [15].

The $Fe_2O_3$ single crystals were cut into pieces of a size of ≈33x25x25 μm³, 14x10x12 μm³ and 13x19x15 μm³. We prepared NaCl plates (size ≈ 120x100 μm²) by compressing NaCl between the diamond anvils to a thickness of ≈ 20 μm. They were subsequently used as a buffer medium to insulate the samples from the diamond anvils below in order to prevent chemical reactions during heating [11]. We placed a ruby sphere next to the NaCl plate for pressure measurements [16,17]. The diamond anvil cell preparation was accomplished by loading pressure liquified argon as pressure transmitting medium in a gas-loading system (Fig.5).

4. Results and Discussion

One of the most difficult problems in melting experiments in DACs is the detection of the exact beginning of a melting event [18]. It is clear and we noticed that in absence of phase transitions and chemical reaction the shape of the solid sample upon heating does not change significantly. However, upon melting the surface tension tries to keep the surface of a liquid phase at minimum. If a certain temperature is achieved, either the whole sample fragment, or at least some part undergoes a drastical change accompanied by rounding of surfaces and the formation of one or more spherical drops (Figs. 3-5). This observation can be clearly associated with the event of liquid-melt formation from the solid sample. The formation of drops in experiments with laser-heated DACs involving melting was reported before, e.g. by Bouhifd & Jephcoat [19], which reported the observation of molten metal drops in element-partitioning experiments in DACs. Molten metallic drops were also observed in DAC experiments on alloying of immiscible elements [20]. However, no systematical use of visual observations of formation of spherical drops for investigation of melting at high pressure in laser-heated DACs has been introduced so far.

We found that for iron and gold a stepwise increase of the laser power at fixed "cold pressure" (i.e. measured at ambient temperature) allows registering temperature of the event of drops formation (i.e. melting) rather precisely (with about 50 K uncertainty between consequent steps in the laser power changes) (Figs. 4,5). In case of hematite, an attempt to heat samples resulted in an immediate raise of temperature to about 2400 K, well above the melting point. Although for hematite we could demonstrate formation of drops upon heating (Fig. 5), we could not measure its melting curve systematically. We noticed also the formation of corona-like rings around hematite crystals (Fig. 5), probably due to the presence of water in rather hygroscopic NaCl used as a thermal insulation or due to chemical reaction between NaCl and $Fe_2O_3$.

The data on melting of gold and iron was collected at several pressures (Fig. 6,7). Figure 6 shows our experimental results for gold compared to melting curves reported in literature [21-23], which originate from experiments either in large volume press [21], or in a piston-cylinder high-pressure apparatus [22,23]. Temperature in these experiments was measured by means of thermocouples and melting was detected by means of differential thermal analysis [22,23] or resistivity measurements [21]. To our knowledge, there is no experimental data on melting of gold loaded into DAC at pressures above 10 GPa, so that a direct comparison between two or more DAC studies is not possible. The equation suggested in Ref. 21 is a result of an experimental work and perfectly describes our experimental data points obtained in this study.

Iron is assumed to be one of the major components in the Earth's core, so that there are many different studies of Fe at non-ambient conditions [3,12,24,25]. In Fig. 7 our data is compared to melting curves available in the literature, namely from the computational study [24], from a combined DAC laser-heating and shock wave study [25], and from two other DAC laser-heating studies [3,12]. Temperature in all referred experiments was measured using the multi-wavelength spectroradiometry technique. The experimental P,T ranges investigated in the referred works overlap and exceed the P,T range covered by our measurements, thus making a comparison straightforward. Williams et. al. [25] reported the annihilation of grain boundaries as a melting criterion, surface-textural features (both after quenching), and the observation of fluid-like migration within the sample while holding high temperature. Böhler et.al. [3] and Shen et.al. [12] employed traceable changes in the X-ray diffraction patterns (described above) as melting criteria. The results based on the X-ray diffraction reveal a tendency to overestimate the melting temperature most likely due to volume effects and thermal gradients inside the sample [3]. However, taking into account the uncertainty in melting temperatures in previously published works [3,12] and our measurements, the data are in good agreement with the melting curves of iron obtained using X-ray diffraction technique.

## 5. Conclusion

In this study we applied a portable laser-heating system to carry out experiments on melting of solids under high pressure in diamond anvil cells. We heated metals (gold and iron) and an oxide (hematite) and observed *in situ* formation of spherical drops from all samples in consequence of the laser heating. Considering the appearance of the spherical drops as a criterion of melting, the melting curves for the three materials under investigation were obtained in good agreement with the data reported in literature for the overlapping pressure range. The melting curve of gold was measured up to 35 GPa that considerably extends experimental information previously available only to 10 GPa.

Acknowledgments:

We thank the German Science Foundation (Deutsche Forschungsgemeinschaft, DFG) for financial support through the project MI605/3-1 within the DFG priority program 1236 (Schwerpunktprogramm, SPP 1236).

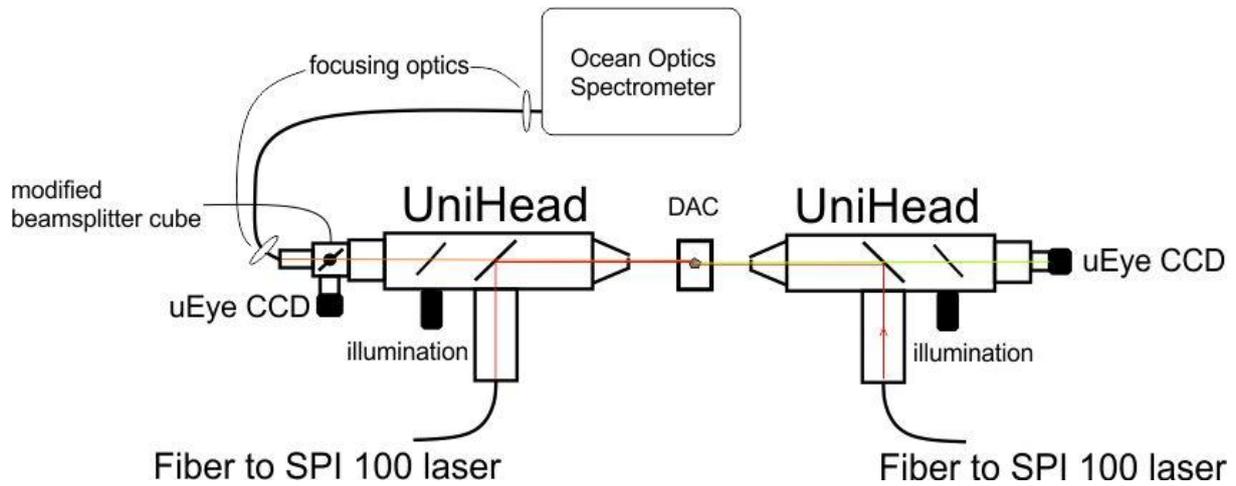

Fig. 1. The schematic experimental setup for laser heating experiments employing two laser heating heads (UniHeads) in opposite geometry; three optical paths: red - 1064 nm laser, orange - thermal emission spectra and green for visualization. The setup consists of laser heating heads (UniHead) with illumination and CCD cameras, two 100 W SPI 100 modulated high power fiber lasers, 532 nm alignment laser, modified visualization and optical output module and QE65000 (Ocean Optics Inc.) multiwavelenght spectrometer.

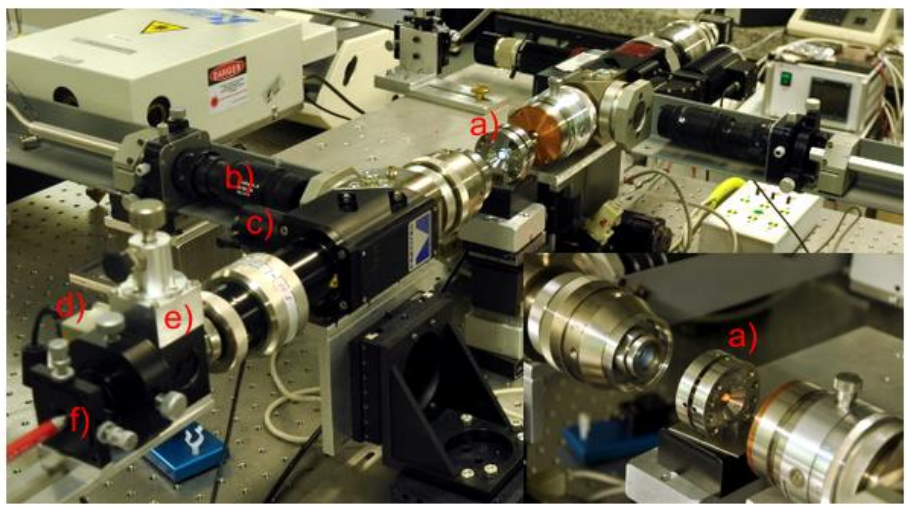

Fig. 2. Photograph of the modified laser heating system (UniHead) (left part of Fig. 1 is described): (a) the position of the DAC, (b) the connector to an SPI 100 fiber laser, (c) the illumination unit for a CCD camera, (d) a μEYE[TM] CCD camera for *in situ* observations, (e)

the movable beam-splitter cube, (f) the fiber to a multiwavelenght spectrometer for temperature measurements (Ocean Optics Inc.). Image detail (lower left part) demonstrates the DAC position between two laser heating heads (UniHeads).

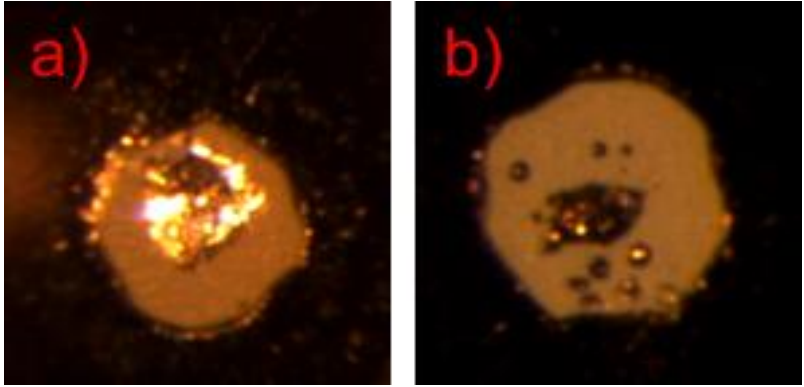

Fig. 3. A sample of iron in a Ne pressure medium after heating at 41(1) GPa and 2520(25) K (a) and after melting at 43(1) GPa and 2560(25) K (b).

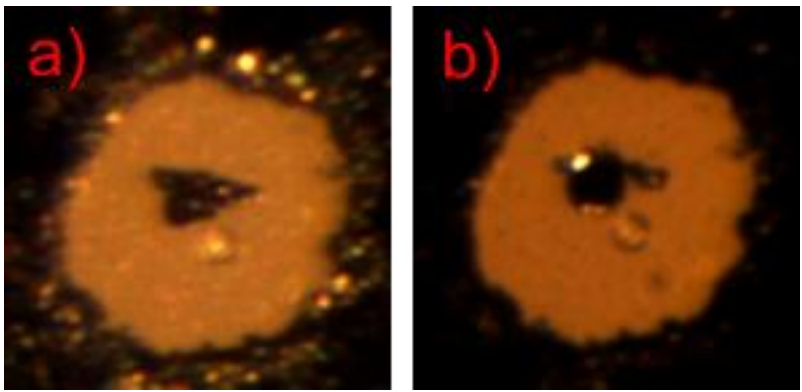

Fig. 4. A sample of gold after heating at 33(1) GPa and 2510 (25) K (a) and after melting at 34(1) GPa and 2590(25) K (b).

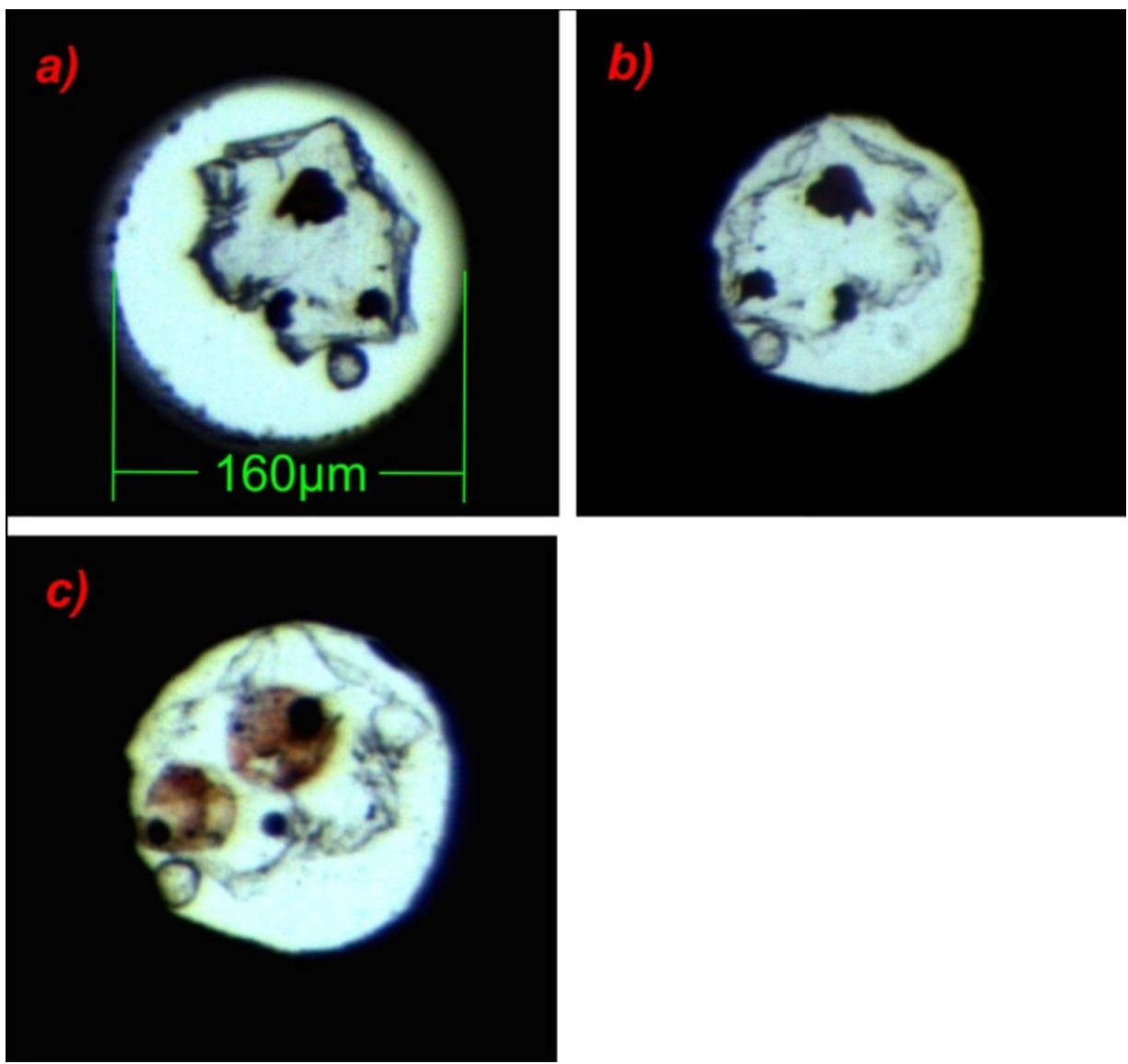

Fig. 5. Photographs of a DAC's content in experiments on melting of hematite: a) the sample assembly after Ar-loading (~ 2 GPa); the central part occupied by the NaCl plate (transparent) with 3 hematite crystals (black) on top; the ruby sphere is placed next to the NaCl plate; b) the same cell pressurized to 10.2 GPa; c) the result of heating sample above 2400(50) K, showing three spheres (black) and two reaction coronas (brown).

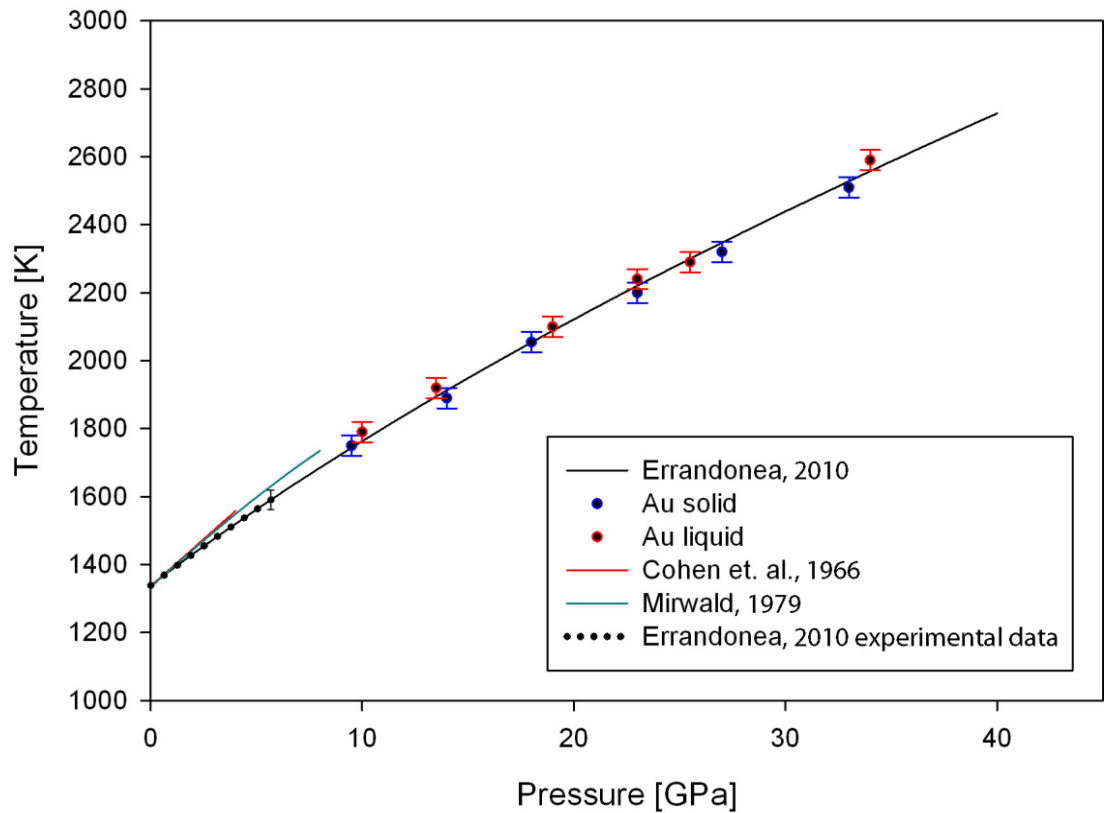

Fig. 6. Melting curves of gold: our present data (blue dots = solid gold, red dots = liquid gold, melting temperature was registered on formation of spherical drops) compared to the melting curves reported in literature (Ref. 21, 22, 23). Error bars, if shown, are taken form original figures.

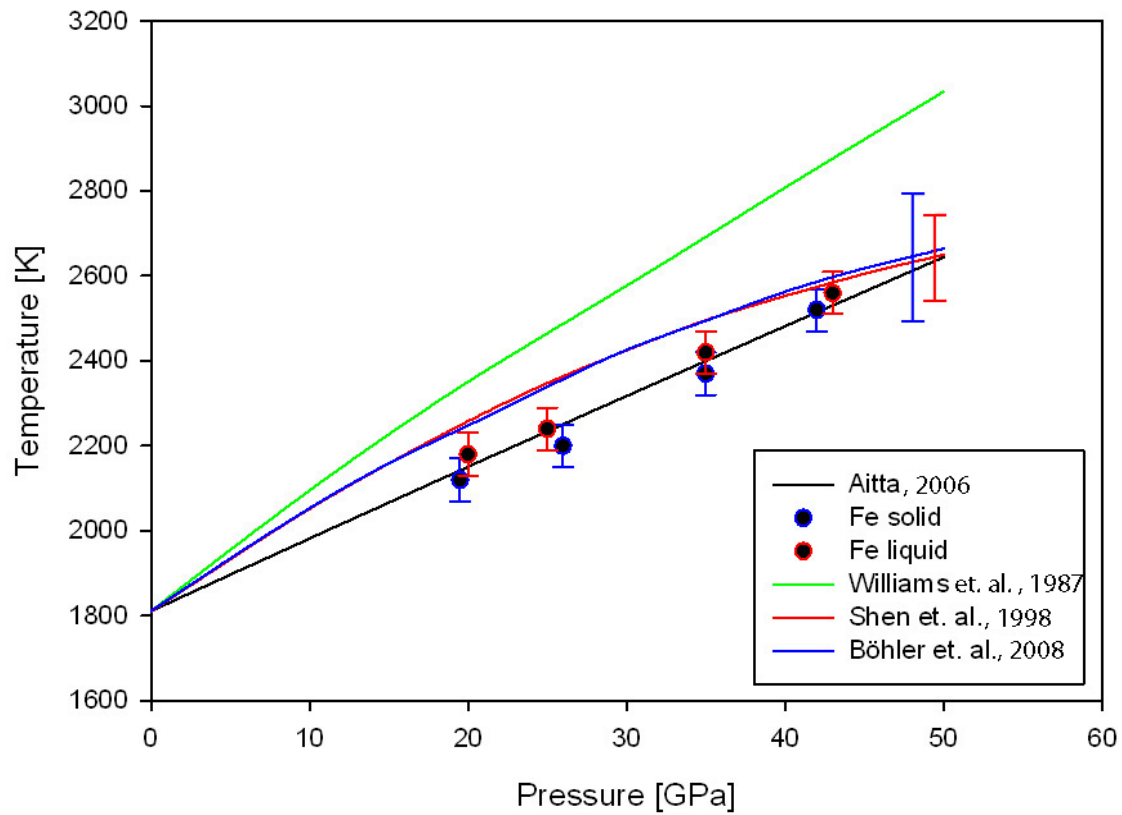

Fig. 7. Melting curves of iron: our observed data (blue dots = solid iron, red dots = liquid iron) compared with the melting curves reported in literature (Ref. 3, 12, 24, 25); error bars are taken, if available, form original figures.